\documentclass[12pt]{revtex4}
\usepackage{dcolumn}
\usepackage{graphicx}
\usepackage{amsmath}
\usepackage{amsfonts}
\usepackage{amssymb}
\usepackage{psfrag}
\usepackage{wrapfig}
\usepackage{subfigure}
\usepackage{makeidx}
\usepackage{bm}
\usepackage{epsf}
\usepackage{psfrag}
\usepackage{hyperref}

\begin{document}

\title{Thermodynamics of third order Lovelock anti-de Sitter black holes revisited}

\author{Decheng Zou$^{1}$,  Ruihong Yue$^{2}$ \footnote{Email:yueruihong@nbu.edu.cn}
and Zhanying Yang$^{1}$\footnote{Email:zyyang@nwu.edu.cn}}
\affiliation{ $^{1}$Department of Physics, Northwest University, Xi'an, 710069, China\\
$^{2}$Faculty of Science, Ningbo University, Ningbo 315211, China}

\baselineskip=20pt

\date{\today}

\begin{abstract}
We compute the mass and the temperature of third order Lovelock black holes with negative
Gauss-Bonnet coefficient $\alpha_2<0$ in anti-de Sitter space and perform the stability analysis
of topological black holes. When $k=-1$, the third
order Lovelock black holes are thermodynamically stable for the whole range $r_+$.
When $k=1$, we found that the black hole has an intermediate unstable phase for $D=7$.
In eight dimensional spacetimes, however, a new phase of thermodynamically unstable small
black holes appears if the coefficient $\tilde{\alpha}$ is under a critical value.
For $D\geq 9$, black holes have similar the distributions of thermodynamically stable regions
to the case where the coefficient $\tilde{\alpha}$ is under a critical value for $D=8$.
It is worth to mention that all the thermodynamic and conserved quantities of the black holes with
flat horizon don't depend on the Lovelock coefficients and are the
same as those of black holes in general gravity.
\end{abstract}



\maketitle

\section{Introduction}
\indent
In the last decade anti-de Sitter (AdS) black holes and especially their thermodynamics
have attracted considerable interest due to the AdS/CFT duality. According to the
AdS/CFT conjecture the thermodynamics of the AdS black holes is related to the
thermodynamics of the dual CFT residing on the boundary of the AdS space. It is well-know that
the AdS Schwarzschild black hole is thermodynamically unstable when the horizon radius is
small, while it is stable for large radius; there is a phase transition, named Hawking-Page
transition\cite{Hawking:1982dh}, between the large stable black hole and a thermal AdS space.
This phase transition is explained by Witten as the confinement/deconfinement transition of
Yang-Mills theory in the AdS-CFT correspondence \cite{Witten:1998zw}.

Besides Einstein-Hilbert action, in their low-energy limit string theories give rise to
effective models of gravity in higher dimensions which involve the higher powers of curvature terms \cite{Lust:1989tj}.
However, the higher powers of curvature could in common give rise to a fourth or even higher order
differential equation for the metric, and it would introduce ghosts and violate unitarity, therefore,
the higher derivative terms may be a source of inconsistencies. Among the gravity theories with
higher derivative terms, the so-called Lovelock gravity \cite{Lovelock:1971yv} is quite special.
Its equations of motion contain the most symmetric conserved tensor with no more than two
derivative of the metric and it has been proven to be free of ghosts when expanding about the
flat space, evading any problem with unitarity \cite{Zwiebach:1985uq}.
In this paper, we restrict ourself to explore the first four terms of the Lovelock
gravity. Among these terms, the first term is cosmological term, the second term is Einstein term,
and the third and the fourth terms are the second order Lovelock(Gauss-Bonnet) and third order
Lovelock terms, respectively.

In third order Lovelock gravity, the static spherically symmetric black hole solutions were firstly
found in \cite{Dehghani:2005zzb}. Then, they also showed that the
asymptotically flat uncharged black hole of third order Lovelock gravity may has two
horizons, a fact that does not happen in lower order Lovelock gravity.
The thermodynamics of the uncharged static black hole solutions with negative cosmological
constant has been considered in \cite{Dehghani:2009zzb}. We note that the Gauss-Bonnet
coefficient always holds on positive in these papers. Here we will study third order Lovelock
black hole solutions with the negative Gauss-Bonnet coefficient in anti-de Sitter space.

The outline of this paper is as follows. Considering the coefficient of Gauss-Bonnet term
$\alpha_2<0$, we present static solution with special values of $\alpha_{2}$ and
$\alpha_{3}$ in section \ref{sec:2}. Then, we discuss some related thermodynamic properties
of black holes in D-dimensional spacetimes. According to the classification of horizon structures,
$k=0$ and $k\pm1$, we will analyze the stability of black holes in section \ref{sec:4}.
Section \ref{sec:3} devotes to conclusions.

\section{Black holes in AdS space}
\label{sec:2}
\indent
The action of third order Lovelock gravity is given by
\begin{eqnarray}
{\cal I}=\frac{1}{16\pi G}\int d^{D}x\sqrt{-g}(R-2\Lambda+\alpha_{2}{\cal L}_{2}+\alpha_{3}{\cal L}_{3})\label{eq:1a},
\end{eqnarray}
where $\Lambda=-\frac{(D-1)(D-2)}{2l^2}$ is a negative cosmological constant, $\alpha_{2}$
and $\alpha_{3}$ are Gauss-Bonnet and third order Lovelock coefficients, respectively.
In Eq.~(\ref{eq:1a}), the Gauss-Bonnet Lagrangian is
\begin{eqnarray}
{\cal L}_{2}=R_{\mu\nu\gamma\delta}R^{\mu\nu\gamma\delta}-4R_{\mu\nu}R^{\mu\nu}+R^2\nonumber
\end{eqnarray}
and the third order Lovelock Lagrangian is
\begin{eqnarray}
{\cal L}_{3}&=&R^3+2R^{\mu\nu\sigma\kappa}R_{\sigma\kappa\rho\tau}R^{\rho\tau}_{~~\mu\nu}
+8R^{\mu\nu}_{~~\sigma\rho}R^{\sigma\kappa}_{~~\nu\tau}R^{\rho\tau}_{~~\mu\kappa}\nonumber\\
&+&24R^{\mu\nu\sigma\kappa}R_{\sigma\kappa\nu\rho}R^{\rho}_{\mu}
+3RR^{\mu\nu\sigma\kappa}R_{\mu\nu\sigma\kappa}\nonumber\\
&+&24R^{\mu\nu\sigma\kappa}R_{\sigma\mu}R_{\kappa\nu}
+16R^{\mu\nu}R_{\nu\sigma}R^{\sigma}_{~\mu}-12RR^{\mu\nu}R_{\mu\nu}.\nonumber
\end{eqnarray}

We assume the metric being of the following form
\begin{eqnarray}
ds^2=-f(r)dt^2+\frac{1}{f(r)}dr^2+r^2h_{ij}dx^idx^j,\label{eq:2a}
\end{eqnarray}
where $h_{ij}dx^idx^j$ represents the metric of a $(d-2)$-dimensional hypersurfaces with
constant curvature scalar $(D-2)(D-3)k$ and volume $\Sigma_{k}$, here k is a constant.
Without loss of the generality, one may take $k=0$ or $\pm1$.

If we choose
\begin{eqnarray}
\alpha_{2}=\frac{\alpha}{(D-3)(D-4)},\quad
\alpha_{3}=\frac{\alpha^2}{72{D-3\choose 4}},\label{eq:3a}
\end{eqnarray}
the uncharged black hole solution in D-dimensions is described by \cite{Dehghani:2005zzb, Dehghani:2009zzb}
\begin{eqnarray}
f(r)=k+\frac{r^2}{\alpha}[1-(1-\frac{3\alpha}{l^2}+\frac{3\alpha m}{r^{D
-1}})^\frac{1}{3}],\label{eq:4a}
\end{eqnarray}

This type of action Eq.~(\ref{eq:1a}) is derived in the low-energy limit of heterotic superstring
theory. Thus, the coefficients $\alpha_{2}$ and $\alpha_{3}$ are regarded as inverse
string tension and positive definite. While, the case for $\alpha_{2}<0$ is also available
to study their black holes. Meanwhile, the corresponding one of the third order Lovelock
term maintains positive. Here, we consider this case $\alpha=-\tilde{\alpha}$
with $\tilde{\alpha}>0$. Therefore, the black hole solution Eq.~(\ref{eq:4a}) becomes
\begin{eqnarray}
f(r)=k-\frac{r^2}{\tilde{\alpha}}[1-(1+\frac{3\tilde{\alpha}}{l^2}-\frac{3\tilde{\alpha} m}
{r^{D-1}})^\frac{1}{3}],\label{eq:5a}
\end{eqnarray}
where the gravitational mass $M$ is expressed as $\frac{(D-2)\Sigma_{k}}{16\pi G} m$.
Since the third order Lovelock term in Eq.~(\ref{eq:1a}) has no contribution to the field
equation in six or less dimensional spacetimes, we consider D-dimensional spacetimes
with $D\geq7$.

On the other hand, the metric Eq.~(\ref{eq:2a}) goes to AdS space asymptotically.
In the limit of $r\rightarrow+\infty$, we only hold back the first two terms of the
taylor expansion $(1+\frac{3\tilde{\alpha}}{l^2}-\frac{3\tilde{\alpha} m}
{r^{D-1}})^\frac{1}{3}$. The asymptotic form for $f(r)$ is expressed as
\begin{eqnarray}
f_{\infty}(r)=k-\frac{r^2}{\tilde{\alpha}}[1-(1+\frac{3\tilde{\alpha}}{l^2})^{1/3}]
-\frac{m}{r^{D-3}(1+\frac{3\tilde{\alpha}}{l^2})^{2/3}}.\label{eq:6a}
\end{eqnarray}

In general relativity, the Schwarzchild-AdS black hole solution in D-dimensional
spacetimes is \cite{Birmingham:1998nr}
\begin{eqnarray}
f(r)=k+\frac{r^2}{l^2}-\frac{m}{r^{D-3}}.\label{eq:7a}
\end{eqnarray}
Hence, we can read off the effective cosmological constant and effective
gravitational mass
\begin{eqnarray}
\frac{1}{l_{eff}^2}=\frac{1}{\tilde{\alpha}}[(1+\frac{3\tilde{\alpha}}{l^2})^{1/3}-1],\quad
M_{eff}=\frac{M}{(1+\frac{3\tilde{\alpha}}{l^2})^{2/3}}.\label{eq:8a}
\end{eqnarray}

Note that the gravitational mass of black hole is determined by $f(r_+)=0$.
From Eq.~(\ref{eq:5a}), the mass can be expressed as in terms of the horizon radius $r_+$
\begin{eqnarray}
M=\frac{(D-2)\Sigma_{k} r_+^{D-3}}{16\pi G}(k+\frac{r_+^2}{l^2}
-\frac{\tilde{\alpha} k^2}{r_+^2}+\frac{\tilde{\alpha}^2k}{3r_+^4}).\label{eq:9a}
\end{eqnarray}
The Hawking temperature  associated with the black hole horizon can be obtained by
requirement of the absence of conical singularity at the horizon in the Euclidean
section of the third order Lovelock black hole solution in AdS spacetimes. According
to $T=\frac{f'(r_+)}{4\pi}$, the temperature of black hole is given by
\begin{eqnarray}
T&=&\frac{1}{12\pi r_+(r_+^2-k\tilde{\alpha})^2}[\frac{3(D-1)r_+^6}{l^2}
+3(D-3)r_+^4k\nonumber\\
&-&3(D-5)r_+^2\tilde{\alpha} k^2+(D-7)\tilde{\alpha}^2k].\label{eq:10a}
\end{eqnarray}

Another important thermodynamic quantity is the entropy of black hole. In general
relativity, the entropy of black hole satisfy the so-called area formula, namely
entropy equals to one quarter of horizon area \cite{Hawking:1974rv}. However, the
area law of entropy is not satisfied in general in higher derivative
gravity \cite{Jacobson:1993xs, Dehghani:2009zzb}. While, as a thermodynamic system,
the black hole must obey the first law of black hole thermodynamics, $dM=TdS$
and then we have
\begin{eqnarray}
S&=&\int_{0}^{r_+} T^{-1}\frac{\partial M}{\partial r_+}dr_+\nonumber\\
&=&\frac{\Sigma_{k}r_+^{D-2}}{4G}[1-\frac{2(D-2)k \tilde{\alpha}}{(D-4)r_+^2}
+\frac{(D-2)k^2\tilde{\alpha}^2}{(D-6)r_+^4}],\label{eq:11a}
\end{eqnarray}
where
\begin{eqnarray}
\frac{\partial M}{\partial r_+}&=&\frac{(D-2)\Sigma_{k} r_+^{D-8}}{48\pi G}
[\frac{3(D-1)r_+^6}{l^2}+3k(D-3)r_+^4\nonumber\\
&-&3(D-5)r_+^2\tilde{\alpha}k^2+(D-7)\tilde{\alpha}^2 k]\nonumber\\
&=&\frac{(D-2)\Sigma_{k} r_+^{D-7}}{4G}(r_+^2-k\tilde{\alpha})^2T.\label{eq:12a}
\end{eqnarray}

Here, we would like to explore the problem of negative entropy. In Gauss-Bonnet
gravity, it has been extensively studies in \cite{Clunan:2004tb}.
The entropy of third order Lovelock black holes Eq.~(\ref{eq:11a})
can be rewritten as
\begin{eqnarray}
S=\frac{\Sigma_{k}r_+^{D-6}}{4G}[r_+^4-\frac{2(D-2)k \tilde{\alpha}r_+^2}{(D-4)}
+\frac{(D-2)k^2\tilde{\alpha}^2}{(D-6)}].\label{eq:13a}
\end{eqnarray}
We here note that the sign of entropy is determined by
\begin{eqnarray}
S_{Pa}=r_+^4-\frac{2(D-2)k \tilde{\alpha}r_+^2}{(D-4)}
+\frac{(D-2)k^2\tilde{\alpha}^2}{(D-6)}.\label{eq:14a}
\end{eqnarray}

Clearly, if $k=1$, $\tilde{\alpha}<0$ or $k=-1$, $\tilde{\alpha}>0$, the
function $S_{Pa}$ is always positive. For $k=1$, $\tilde{\alpha}>0$ or $k=-1$,
$\tilde{\alpha}<0$, we obtain
\begin{eqnarray}
S_{Pa}=r_+^4-\frac{2(D-2)\tilde{\alpha}r_+^2}{(D-4)}
+\frac{(D-2)\tilde{\alpha}^2}{(D-6)}.\nonumber
\end{eqnarray}
It is interesting to mention that the function $S_{Pa}$ is also positive for $r_+>0$.
As a result, unlike the entropy in Gauss-Bonnet gravity, the entropy of black holes with
special coefficient is always positive in third order Lovelock gravity.

\section{Stability of topological black holes}
\label{sec:3}
\indent
In this section, we perform the stability analysis of topological black holes.
The local thermodynamic stability of black hole is determined by
the sign of its heat capacity. If the heat capacity is positive, we have that
the black hole is locally stable to thermal fluctuations. When the heat capacity
is negative, the black hole is said to be locally unstable.

The heat capacity of black holes is
\begin{eqnarray}
C&=&\frac{\partial M}{\partial T}=(\frac{\partial M}{\partial r_+})
(\frac{\partial r_+}{\partial T}),\label{eq:15a}
\end{eqnarray}
where
\begin{eqnarray}
\frac{\partial T}{\partial r_+}&=&\frac{1}{12\pi l^2r_+^2(r_+^2
-k\tilde{\alpha})^3}[3(D-1)r_+^8-3k(5(D-1)\tilde{\alpha}
+(D-3)l^2)r_+^6\nonumber\\
&-&18\tilde{\alpha} k^2r_+^4-2k(D-10)\tilde{\alpha}^2r_+^2
+(D-7)\tilde{\alpha}^3k^2].\label{eq:16a}
\end{eqnarray}
Then the expression of heat capacity indicating the local stability of the
black hole is obtained
\begin{eqnarray}
C&=&\frac{(D-2)\Sigma_{k} r_+^{D-6}(r_+^2-k\tilde{\alpha})^3 J(r)}{4G\Gamma(r)}\nonumber\\
&=&\frac{3(D-2)\pi\Sigma_{k}l^2r_+^{D-5}(r_+^2-k\tilde{\alpha})^5 T/G}{\Gamma(r)},\label{eq:17a}
\end{eqnarray}
where $J(r)$ and $\Gamma(r)$ are expressed as $3(D-1)r_+^6+3k(D-3)r_+^4 l^2
-3(D-5)r_+^2\tilde{\alpha}k^2 l^2+(D-7)\tilde{\alpha}^2k l^2$ and $3(D-1)r_+^8
-3k(5(D-1)\tilde{\alpha}+(D-3)l^2)r_+^6-18\tilde{\alpha} k^2r_+^4l^2
-2k(D-10)\tilde{\alpha}^2r_+^2l^2+(D-7)\tilde{\alpha}^3k^2l^2$, respectively.

It is clear that those physical properties depend on the horizon structure $k$ and the
dimensions of spacetime. Below, we will discuss each case according to the classification
of horizon structure $k=0$ and $k=\pm1$, respectively.

\subsection{Flat black hole}
\indent
In case of $k=0$, we have
\begin{eqnarray}
&&M=\frac{(D-2)\Sigma_{k}r_+^{D-1}}{16\pi Gl^2},\quad T=\frac{(D-1)r_+}{4\pi l^2},
S=\frac{\Sigma_{k} r_+^{D-2}}{4G},\nonumber\\
&&C=\frac{(D-2)\Sigma_{k} r_+^{D-2}}{4G},\quad F=-\frac{\Sigma_{k} r_+^{D-1}}{16\pi G l^2}.\label{eq:18a}
\end{eqnarray}
One can find that these thermodynamic quantities are independent of the the Lovelock
coefficients, and have the completely same expressions as those in \cite{Birmingham:1998nr}.
We conclude that the higher order derivative terms do not affect the thermodynamic
properties of black holes although they have different black hole solutions.

\subsection{Hyperbolic black hole}
\indent
Now, we turn to the case the horizon is a negative constant curvature hypersurface.
For $D=7$, the temperature of black holes Eq.~(\ref{eq:10a}) reduces
to a simple form
\begin{eqnarray}
T=\frac{r_+}{2\pi(r_+^2+\tilde{\alpha})^2}(3r_+^4/l^2-2r_+^2-\tilde{\alpha}).\label{eq:19a}
\end{eqnarray}
The existence of extremal black holes depend on the existence of positive roots for temperature $T=0$,
which reduces to
\begin{eqnarray}
r_{+}(3r_{+}^4/l^2-2r_{+}^2-\tilde{\alpha})=0.\label{eq:20a}
\end{eqnarray}
Then, the largest positive root of this equation which corresponds to
radius of extremal black hole is obtained
\begin{eqnarray}
r_{ext}=l[(1+\sqrt{1+3\tilde{\alpha}/l^2})/3]^{1/2}.\label{eq:21a}
\end{eqnarray}
The black holes solution presents a black hole provided $r_+>r_{ext}$.

Substituting $r_{ext}$ into the gravitational mass Eq.~(\ref{eq:9a}), the mass of extremal
black hole is given by
\begin{eqnarray}
M_{ext}&=&\frac{(D-2)\Sigma_k r_{ext}^4}{16\pi G}(\frac{r_{ext}^2}{l^2}-1
-\frac{\tilde{\alpha}}{r_{ext}^2}-\frac{\tilde{\alpha}^2}{3r_{ext}^4})\nonumber\\
&=&\frac{-(D-2)l^8\Sigma_k}{16\pi G}\frac{(3\tilde{\alpha}/l^2+1)(1+\sqrt{1
+3\tilde{\alpha}/l^2})^4}{3^5}.\nonumber
\end{eqnarray}

In Fig.~1, we show the temperature $T$ with $\tilde{\alpha}=0.1$ and 0.2 in seven
dimensional spacetimes, respectively. The temperature $T$ always
starts from zero at $r_+=r_{ext}$ and then goes to positive infinity as $r_+$ increases.
Obviously, $\frac{\partial T_H}{\partial r_H}$ Eq.~(\ref{eq:16a}) is positive for $r_+>r_{ext}$.
The heat capacity $C$ Eq.~(\ref{eq:15a}) is the
product of $\frac{\partial M}{\partial r_H}$ and $\frac{\partial r_H}{\partial T_H}$.
We note that $\frac{\partial M}{\partial r_H}$ Eq.~(\ref{eq:12a}) is always positive
for $T>0$. In the $T=0$ case, one can see from Eq.~(\ref{eq:16a}) that besides $r_+=0$, the heat
capacity $C$ also vanishes at $r_+=r_{ext}$. In Fig.~2, we show the graph of heat
capacity $C$ versus horizon radius $r_+$ and the larger root of $C$ corresponds
to $r_{ext}$. Therefore, the heat capacity $C$ is always positive and then black holes
are thermodynamically stable.

%%%%%%%%%%%%%%%%%%%%%%%%%%%%%%%%%%%%%%%%%%%%%%%%%%%%%%%%%%%%%%%%%%%%%%%%%%%
\begin{figure}[htb]
\begin{minipage}[t]{0.43\linewidth}
\centering
\includegraphics{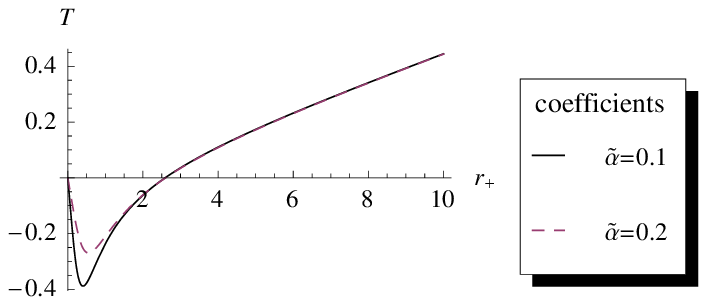}
\caption{Temperature $T$ versus horizon radius $r_+$ with $k=-1$, $l^2=10$ and $D=7$.}
\end{minipage}%
\hfill%
\begin{minipage}[t]{0.43\linewidth}
\centering
\includegraphics{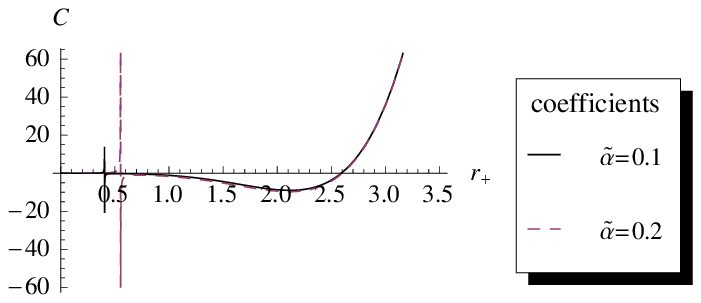}
\caption{Heat capacity $C$ versus horizon radius $r_+$ with $k=-1$, $l^2=10$ and $D=7$.}
\end{minipage}
\end{figure}

We can easily extend all of the discussions of the previous subsubsection to $D$-dimensional
solutions. When $D\geq 8$, the existence of extremal black hole depends on equation $T=0$
\begin{widetext}
\begin{eqnarray}
\frac{3(D-1)r_{ext}^6}{l^2}-3(D-3)r_{ext}^4-3(D-5)r_{ext}^2\tilde{\alpha}
-(D-7)\tilde{\alpha}^2=0.\label{eq:22a}
\end{eqnarray}
If ordering $r_{ext}^2=R_{ext}>0$, we have a cubic equation
\begin{eqnarray}
R_{ext}^3-\frac{(D-3)l^2}{(D-1)}R_{ext}^2-\frac{(D-5)l^2}{(D-1)}\tilde{\alpha}R_{ext}
-\frac{(D-7)l^2}{3(D-1)}\tilde{\alpha}^2=0.\label{eq:23a}
\end{eqnarray}

In order to find the exact solution, let
\begin{eqnarray}
a_{1}=-\frac{(D-3)l^2}{(D-1)},\quad a_{2}=-\frac{(D-5)l^2}{(D-1)}\tilde{\alpha},\quad
a_{3}=-\frac{(D-7)l^2}{3(D-1)}\tilde{\alpha}^2\label{eq:24a}
\end{eqnarray}
and we have
\begin{eqnarray}
Q&=&\frac{3a_{2}-a_{1}^2}{9}=-\frac{l^2[3\tilde{\alpha}(D-1)(D-5)+(D-3)^2l^2]}{9(D-1)^2}\nonumber\\ P&=&\frac{9a_{1}a_{2}-27a_{3}-2a_{1}^3}{54}\nonumber\\
&=&l^2[\frac{\tilde{\alpha}^2(D-7)}{6(D-1)}
+\frac{\tilde{\alpha}(D-3)(D-5)l^2}{6(D-1)^2}+\frac{(D-3)^3l^4}{27(D-1)^3}].\label{eq:25a}
\end{eqnarray}
Then, we obtain the discriminant of this cubic equation $\Delta=Q^3+P^2$
\begin{eqnarray}
\Delta&=&\frac{\tilde{\alpha}^2(D-1)}{4l^4}[9(D-7)^2(D-1)\tilde{\alpha}^2+6\tilde{\alpha}(D-5)(13
-10D+D^2)l^2\nonumber\\
&+&(D-9)(D-3)^2l^4]\nonumber\\
&=&\frac{\tilde{\alpha}^2(D-1)}{4l^4}[\tilde{\alpha}+\frac{l^2}{3}][\tilde{\alpha}
+\frac{(D-9)(D-3)^2l^2}{3(D-1)(D-7)^2}].\label{eq:26a}
\end{eqnarray}
\end{widetext}

Depending on the choice of the parameter $\tilde{\alpha}$, the discriminant $\Delta$
has different signs and disappears at
\begin{eqnarray}
\tilde{\alpha}^{(1)}=-\frac{l^2}{3},\qquad
\tilde{\alpha}^{(2)}=-\frac{(D-9)(D-3)^2l^2}{3(D-1)(D-7)^2}.\label{eq:27a}
\end{eqnarray}
For $\Delta \geq 0$, the solution of the cubic equation Eq.~(\ref{eq:23a})
can be written down as
\begin{displaymath}
\left\{ \begin{array}{ll}
R_1=S+T-\frac{a_1}{3}, \\
R_2=-\frac{1}{2}(S+T)-\frac{a_1}{3}+\frac{1}{2}i\sqrt{3}(S-T), \\
R_3=-\frac{1}{2}(S+T)-\frac{a_1}{3}-\frac{1}{2}i\sqrt{3}(S-T),
\end{array} \right.
\end{displaymath}
where $S=\sqrt[3]{P+\sqrt{\Delta}}$ and $T=\sqrt[3]{P-\sqrt{\Delta}}$.
If $\Delta < 0$, the solution is
\begin{displaymath}
\left\{\begin{array}{ll}
\tilde{R}_1=2\sqrt{-Q}\cos(\theta/3)-\frac{a_1}{3}, \\
\tilde{R}_2=2\sqrt{-Q}[\cos(\theta/3)+120^\circ]-\frac{a_1}{3}, \\
\tilde{R}_3=2\sqrt{-Q}[\cos(\theta/3)+240^\circ]-\frac{a_1}{3},
\end{array} \right.
\end{displaymath}
where $\theta=\arccos P/\sqrt{-Q^3}$.

For $D=8$, $\tilde{\alpha}^{(2)}$ is positive and so there exist two case
$0<\tilde{\alpha}<\tilde{\alpha}^{(2)}$ and $\tilde{\alpha}>\tilde{\alpha}^{(2)}$.
As to the latter case $\tilde{\alpha}>\tilde{\alpha}^{(2)}$, we have only one positive
real root $r_{ext}=\sqrt{R_1}$ of the cubic equation Eq.~(\ref{eq:23a}).
For $0<\tilde{\alpha}<\tilde{\alpha}^{(2)}$, the discriminant $\Delta$ is negative and
then the cubic equation has one positive $r_{ext}=\sqrt{\tilde{R}_1}$ and
two others negative roots. In case of $D\geq9$, $\tilde{\alpha}^{(2)}$ is non-positive. Then, the
discriminant $\Delta$ is positive for arbitrary value of coefficient $\tilde{\alpha}$
and the cubic equation only has one positive root $r_{ext}=\sqrt{R_1}$. Hence,
in eight or higher dimensional spacetimes, the cubic equation has a positive root which
corresponding to the radius of extremal black hole and the temperature $T$ only vanishes
at $r_+=r_{ext}$. In Fig.~3, the temperature $T$ is shown for $D=8$ and 9. Taking $l^2=10$
and  $D=8$, we have $\tilde{\alpha}^{(2)}\approx11.90$ and $r_{ext}\approx2.70$. The graph
of heat capacity $C$ is plotted with different parameters $\tilde{\alpha}$ in Fig.~4.
Since the equations $\frac{\partial r_H}{\partial T_H}$ and
$\frac{\partial M}{\partial r_H}$ always maintain positive for $T>0$,
the heat capacity $C$ is always positive for the
temperature $T>0$ and the black holes are thermodynamically stable.

%%%%%%%%%%%%%%%%%%%%%%%%%%%%%%%%%%%%%%%%%%%%%%%%%%%%%%%%%%%%%%%%%%%%%%%%%%%
\begin{figure}[htb]
\centering
\subfigure[$\tilde{\alpha}=0.1<\tilde{\alpha}^{(2)}$]{
\label{fig:subfig:a} %% label for first subfigure
\includegraphics{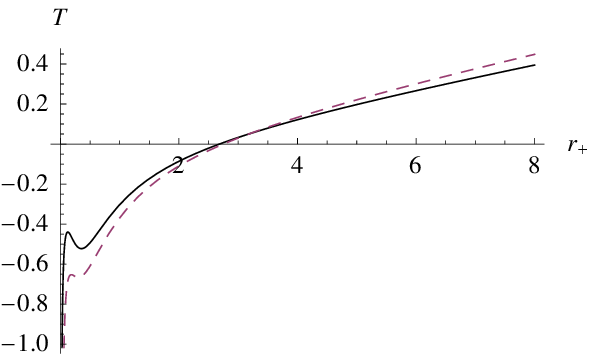}}%
\hfill%
\subfigure[$\tilde{\alpha}=12>\tilde{\alpha}^{(2)}$]{
\label{fig:subfig:b} %% label for second subfigure
\includegraphics{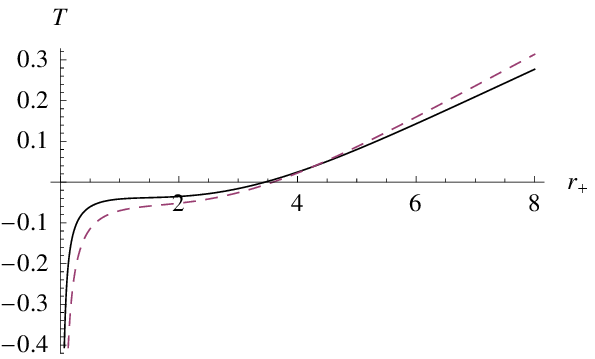}}
\caption{ Temperature $T$ versus horizon radius $r_+$ for $k=-1$, $l^2=10$, $D=8$ (solid line)
and $D=9$ (dashed line).}
\label{fig:subfig} %% label for entire figure
\end{figure}

\begin{figure}[htb]
\centering
\subfigure[$\tilde{\alpha}=0.1<\tilde{\alpha}^{(2)}$]{
\label{fig:subfig:a} %% label for first subfigure
\includegraphics{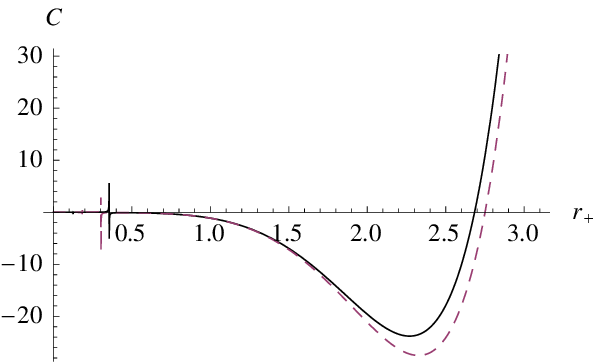}}%
\hfill%
\subfigure[$\tilde{\alpha}=12>\tilde{\alpha}^{(2)}$]{
\label{fig:subfig:b} %% label for second subfigure
\includegraphics{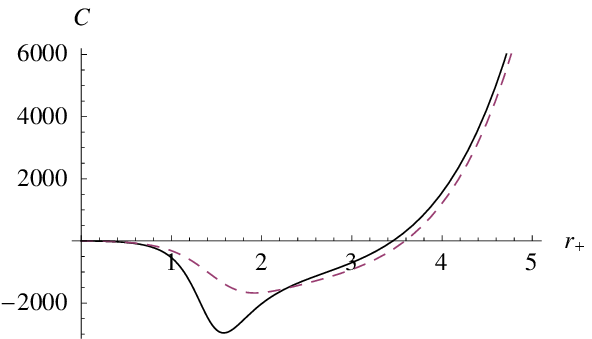}}
\caption{ Heat capacity $C$ versus horizon radius $r_+$ for $k=-1$, $l^2=10$, $D=8$ (solid line)
and $D=9$ (dashed line).}
\label{fig:subfig} %% label for entire figure
\end{figure}

\subsection{Spherical black hole}
\indent
In this subsection we shall explore some physical aspects of the black holes with
positive constant curvature hypersurface horizon.
From Eq.~(\ref{eq:8a}), one can see that there maybe exist a extremal black hole for $T=0$, which
is expressed as
\begin{eqnarray}
r_{+}(3r_{+}^4/l^2+2r_{+}^2-\tilde{\alpha})=0,\label{eq:28a}
\end{eqnarray}
We notice that temperature $T$ has a singularity at $r_{+}=\sqrt{\tilde{\alpha}}$. Then,
largest root which denotes the horizon radius of the extremal black hole is obtain
\begin{eqnarray}
r_{ext}=l[(-1+\sqrt{1+3\tilde{\alpha}/l^2})/3]^{1/2}.\label{eq:29a}
\end{eqnarray}
Indeed the black hole solutions present a black hole
for $r_+>r_{ext}$, an extremal black hole if $r_+=r_{ext}$, and a naked singularity
for $r_+<r_{ext}$. Hence, the gravitational mass of extremal black hole is given by
\begin{eqnarray}
M_{ext}&=&\frac{(D-2)\Sigma_{k} r_{ext}^4}{16\pi G}(1+\frac{r_{ext}^2}{l^2}
-\frac{\tilde{\alpha}}{r_{ext}^2}+\frac{\tilde{\alpha}^2}{3r_{ext}^4})\nonumber\\
&=&\frac{(D-2)\Sigma_{k}}{3^5\times16\pi G}(27\tilde{\alpha}^3+9\tilde{\alpha}^2l^2
+96\tilde{\alpha}l^4+40l^6\nonumber\\
&+&36l^4\tilde{\alpha}\sqrt{1+3\tilde{\alpha}}
+40l^6\sqrt{1+3\tilde{\alpha}}).\label{eq:30a}
\end{eqnarray}

%%%%%%%%%%%%%%%%%%%%%%%%%%%%%%%%%%%%%%%%%%%%%%%%%%%%%%%%%%%%%%%%%%%%%%%%%%%
\begin{figure}[htb]
\centering
\subfigure[$\tilde{\alpha}=0.1$]{
\label{fig:subfig:a} %% label for first subfigure
\includegraphics{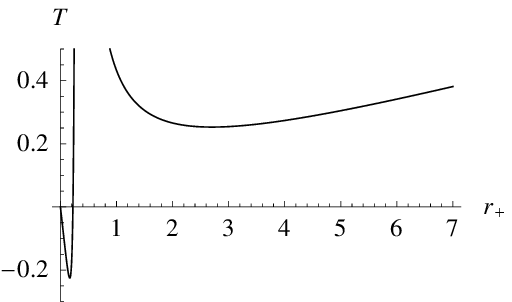}}%
\hfill%
\subfigure[$\tilde{\alpha}=0.2$]{
\label{fig:subfig:b} %% label for second subfigure
\includegraphics{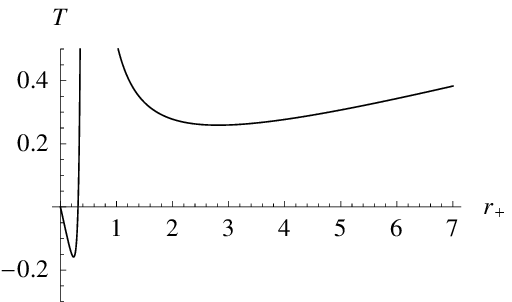}}
\caption{ Temperature $T$ versus horizon radius $r_+$ with $k=1$, $l^2=10$ and $D=7$.}
\end{figure}

\begin{figure}[htb]
\centering
\subfigure[$0.2<r_+<0.6$]{
\label{fig:subfig:a} %% label for first subfigure
\includegraphics{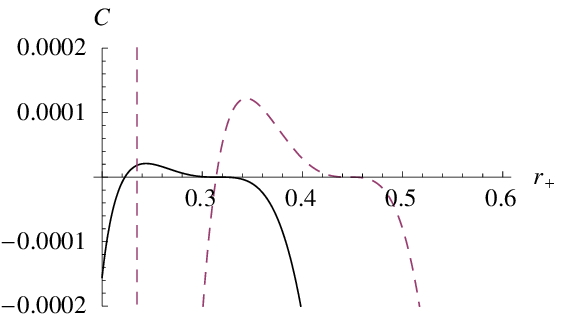}}%
\hfill%
\subfigure[$0.6<r_+<3.9$]{
\label{fig:subfig:b} %% label for second subfigure
\includegraphics{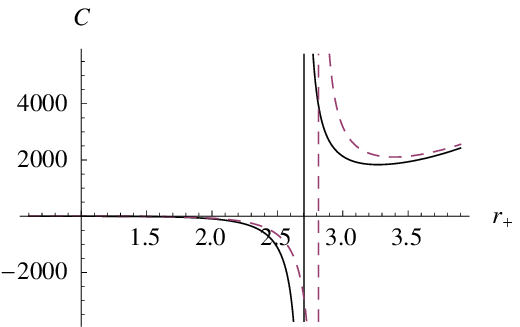}}
\caption{ Heat capacity $C$ versus horizon radius $r_+$ with $k=1$, $l^2=10$, $D=7$,
$\tilde{\alpha}=0.1$ (solid line) and $\tilde{\alpha}=0.2$ (dashed line).}
\label{fig:subfig} %% label for entire figure
\end{figure}

In Fig.~5, we show the temperature $T$ versus horizon radius $r_+$ with different
parameters $\tilde{\alpha}$ in seven dimensional spacetimes.
For $\tilde{\alpha}=0.1$, we can obtain the horizon radius $r_{ext}\approx0.22$.
In Fig.6, heat capacity $C$ is plotted with $\tilde{\alpha}=0.1$. The
heat capacity $C$ vanishes at $r_{ext}\approx0.22$ which corresponding to $T=0$
and $\sqrt{\tilde{\alpha}}\approx0.32$.
Later, it blows up at $r_{c}\approx2.70$ and then changes sign so that $C$
becomes positive. Finally, $C$ gradually goes to positive infinity
as $r_+\rightarrow\infty$. Therefore, the black holes in the
regions $r_{ext}<r_+<\sqrt{\tilde{\alpha}}$ and $r_+>r_{c}$ are locally stable.
For $\sqrt{\tilde{\alpha}}<r_+<r_{c}$, black holes are locally unstable.

For $D\geq 8$, the existence of extremal black hole is determined by
\begin{eqnarray}
0=3(D-1)r_{+}^6/l^2+3(D-3)r_{+}^4-3(D-5)\tilde{\alpha}r_{+}^2
+(D-7)\tilde{\alpha}^2.\label{eq:31a}
\end{eqnarray}
Here we also rewrite it to the form
\begin{eqnarray}
R_+^3+b_1R_+^2+b_2R_++b_3=0,\label{eq:32a}
\end{eqnarray}
where $R_+=r_+^2$ and
\begin{widetext}
\begin{eqnarray}
b_1=\frac{(D-3)l^2}{(D-1)},\quad
b_2=-\frac{(D-5)l^2}{(D-1)}\tilde{\alpha},\quad
b_3=\frac{(D-7)l^2}{3(D-1)}\tilde{\alpha}^2.\label{eq:33a}
\end{eqnarray}
Adopted the same approach above, we have
\begin{eqnarray}
Q&=&\frac{3b_{2}-b_{1}^2}{9}=-\frac{l^2[3\tilde{\alpha}(D-1)(D-5)+(D-3)^2l^2]}{9(D-1)^2}\nonumber\\ P&=&\frac{9b_{1}b_{2}-27b_{3}-2b_{1}^3}{54}\nonumber\\
&=&-l^2[\frac{\tilde{\alpha}^2(D-7)}{6(D-1)}
+\frac{\tilde{\alpha}(D-3)(D-5)l^2}{6(D-1)^2}+\frac{(D-3)^3l^4}{27(D-1)^3}].\label{eq:34a}
\end{eqnarray}
Thus, we can obtain the discriminant of this cubic equation $\Delta=Q^3+P^2$
\begin{eqnarray}
\Delta&=&\frac{\tilde{\alpha}^2(D-1)}{4l^4}[9(D-7)^2(D-1)\tilde{\alpha}^2+6\tilde{\alpha}(D-5)(13
-10D+D^2)l^2+(D-9)(D-3)^2l^4]\nonumber\\
&=&\frac{\tilde{\alpha}^2(D-1)}{4l^4}[\tilde{\alpha}+\frac{l^2}{3}][\tilde{\alpha}
+\frac{(D-9)(D-3)^2l^2}{3(D-1)(D-7)^2}].\label{eq:35a}
\end{eqnarray}
\end{widetext}
We find that it is the same as Eq.~(\ref{eq:26a}). Therefore, the discriminant $\Delta$
vanishes at
\begin{eqnarray}
\tilde{\alpha}^{(1)}=-\frac{l^2}{3},\qquad
\tilde{\alpha}^{(2)}=-\frac{(D-9)(D-3)^2l^2}{3(D-1)(D-7)^2}.\label{eq:36a}
\end{eqnarray}

For $\Delta \geq 0$, the solution of the cubic equation Eq.~(\ref{eq:23a})
can be written down as
\begin{displaymath}
\left\{ \begin{array}{ll}
R_1=S+T-\frac{b_1}{3}, \\
R_2=-\frac{1}{2}(S+T)-\frac{a_1}{3}+\frac{1}{2}i\sqrt{3}(S-T), \\
R_3=-\frac{1}{2}(S+T)-\frac{a_1}{3}-\frac{1}{2}i\sqrt{3}(S-T),
\end{array} \right.
\end{displaymath}
where $S=\sqrt[3]{P+\sqrt{\Delta}}$ and $T=\sqrt[3]{P-\sqrt{\Delta}}$.
If $\Delta < 0$, the solution is
\begin{displaymath}
\left\{\begin{array}{ll}
\tilde{R}_1=2\sqrt{-Q}\cos(\theta/3)-\frac{b_1}{3}, \\
\tilde{R}_2=2\sqrt{-Q}[\cos(\theta/3)+120^\circ]-\frac{b_1}{3}, \\
\tilde{R}_3=2\sqrt{-Q}[\cos(\theta/3)+240^\circ]-\frac{b_1}{3},
\end{array} \right.
\end{displaymath}
where $\theta=\arccos P/\sqrt{-Q^3}$.

For $D=8$, there are two case $0<\tilde{\alpha}<\tilde{\alpha}^{(2)}$ and
$\tilde{\alpha}>\tilde{\alpha}^{(2)}$. For $0<\tilde{\alpha}<\tilde{\alpha}^{(2)}$,
the discriminant $\Delta$ is negative and then this cubic equation has three roots:
two positive $r_>=\sqrt{\tilde{R}_1}$, $r_<=\sqrt{\tilde{R}_2}$ and one
negative $\sqrt{\tilde{R}_3}$. The larger positive one $r_>$ corresponds to the horizon
radius of extremal black hole $r_{ext}$. However, for $\tilde{\alpha}>\tilde{\alpha}^{(2)}$,
the discriminant $\Delta$ is positive and we also have a real which is negative.
It is suggested that the temperature $T$ is always positive for the whole range $r_+\geq 0$
and there is no extremal black hole. In case of $D\geq9$, the discriminant $\Delta$
is always positive. We find that this real one is negative and the temperature of
black hole is always positive for $D\geq9$.
Hence, the extremal black holes only exist for $0<\tilde{\alpha}<\tilde{\alpha}^{(2)}$
in eight dimensional spacetimes. In Fig.7, we plot the temperature versus horizon
radius $r_+$ for $D=8$, 9 and 10, respectively and the details are shown in Fig.~8.
For $l^2=10$ and $D=8$, we have $\tilde{\alpha}^{(2)}=11.90$.

%%%%%%%%%%%%%%%%%%%%%%%%%%%%%%%%%%%%%%%%%%%%%%%%%%%%%%%%%%%%%%%%%%%%%%%%%%%
\begin{figure}[htb]
\centering
\subfigure[$\tilde{\alpha}=0.1$]{
\label{fig:subfig:a} %% label for second subfigure
\includegraphics{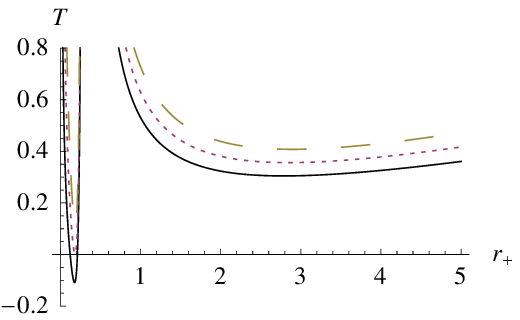}}
\hfill%
\subfigure[$\tilde{\alpha}=12$]{
\label{fig:subfig:b} %% label for second subfigure
\includegraphics{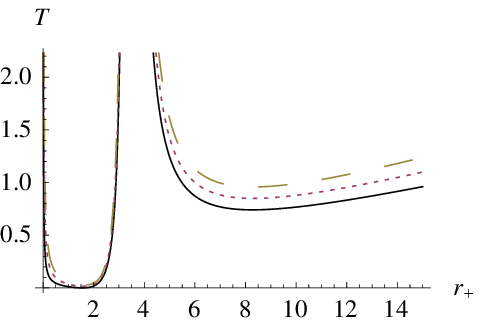}}
\caption{Temperature $T$ versus horizon radius $r_+$ with $l^2=10$ and $k=1$.
These three curves from up to down correspond to $D=10$, $D=9$ and $D=8$, respectively.}
\label{fig:subfig} %% label for entire figure
\end{figure}

%%%%%%%%%%%%%%%%%%%%%%%%%%%%%%%%%%%%%%%%%%%%%%%%%%%%%%%%%%%%%%%%%%%%%%%%%%%
\begin{figure}[htb]
\centering
\subfigure[$\tilde{\alpha}=0.1$, $D=9$]{
\label{fig:subfig:a} %% label for second subfigure
\includegraphics{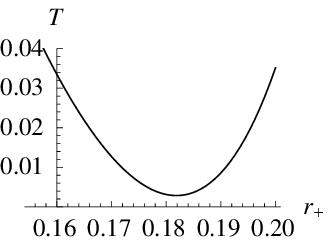}}%
\hfill%
\subfigure[$\tilde{\alpha}=12$, $D=8$]{
\label{fig:subfig:b} %% label for second subfigure
\includegraphics{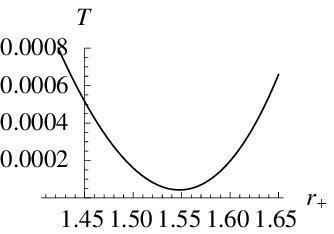}}%
\hfill%
\subfigure[$\tilde{\alpha}=12$, $D=9$]{
\label{fig:subfig:c} %% label for second subfigure
\includegraphics{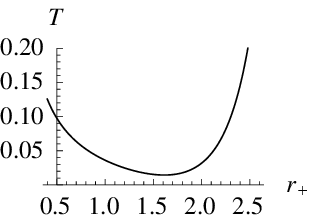}}%
\hfill%
\subfigure[$\tilde{\alpha}=12$, $D=10$]{
\label{fig:subfig:d} %% label for second subfigure
\includegraphics{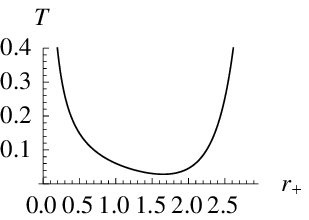}}
\caption{ Temperature $T$ versus horizon radius $r_+$ with $l^2=10$ and $k=1$. }
\label{fig:subfig} %% label for entire figure
\end{figure}

%%%%%%%%%%%%%%%%%%%%%%%%%%%%%%%%%%%%%%%%%%%%%%%%%%%%%%%%%%%%%%%%%%%%%%%%%%%
\begin{figure}[htb]
\centering
\subfigure[$0<r_+<0.39$]{
\label{fig:subfig:a} %% label for second subfigure
\includegraphics{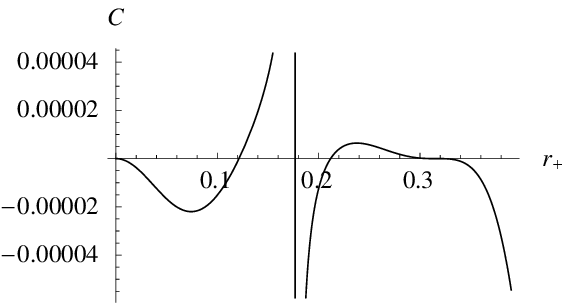}}%
\hfill%
\subfigure[$0.39<r_+<3.4$]{
\label{fig:subfig:b} %% label for second subfigure
\includegraphics{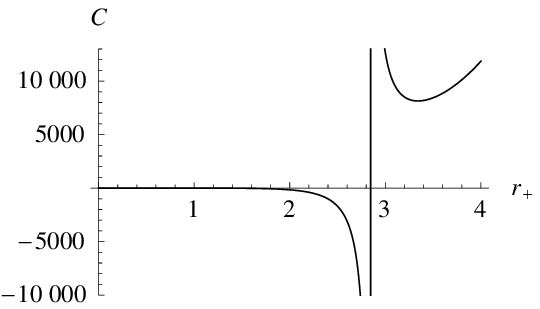}}
\caption{Heat capacity $C$ versus horizon radius $r_+$ with $\tilde{\alpha}=0.1$,
$l^2=10$ and $D=8$.}
\label{fig:subfig} %% label for entire figure
\end{figure}

%%%%%%%%%%%%%%%%%%%%%%%%%%%%%%%%%%%%%%%%%%%%%%%%%%%%%%%%%%%%%%%%%%%%%%%%%%%
\begin{figure}[htb]
\centering
\subfigure[$0<r_+<5$]{
\label{fig:subfig:a} %% label for second subfigure
\includegraphics{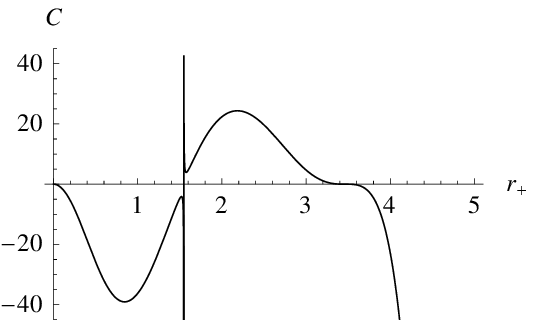}}
\hfill%
\subfigure[$5<r_+<11$]{
\label{fig:subfig:b} %% label for second subfigure
\includegraphics{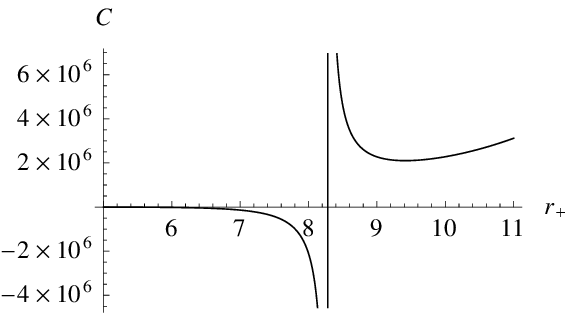}}
\caption{Heat capacity $C$ versus horizon radius $r_+$ with $\tilde{\alpha}=12$, $l^2=10$ and $D=8$.}
\label{fig:subfig} %% label for entire figure
\end{figure}

Here, let us perform the stability analysis of black holes in eight dimensional spacetimes.
First we consider the black holes with coefficient $\tilde{\alpha}<\tilde{\alpha}^{(2)}$.
The heat capacity $C$ Eq.~(\ref{eq:17a}) disappears at $r_+=r_{ext}$ and $r_+=\sqrt{\tilde{\alpha}}$.
In Fig.9, we show heat capacity
$C$ versus horizon radius $r_+$ with $\sqrt{\tilde{\alpha}}=0.1$. It shows that these two
larger roots correspond to extremal horizon $r_{ext}$ and $\sqrt{\tilde{\alpha}}$ and $C$
is always positive in the region $r_{ext}<r_+<\sqrt{\tilde{\alpha}}$ and the black holes
are always local stable. If taking $l^2=10$ and $\tilde{\alpha}=0.1$, we can obtain the
extremal radius $r_{ext}\approx0.22$. For $r_+>\sqrt{\tilde{\alpha}}$, the heat capacity $C$
becomes negative in the region $r_+>\sqrt{\tilde{\alpha}}$
and gradually blows up at $r_+=r_{cs}$ (say) and then changes sign becomes positive. Hence,
black holes are local unstable in the domain $\sqrt{\tilde{\alpha}}<r_+<r_{cs}$ and are stable
for $r_+>r_{cs}$. For $\tilde{\alpha}>\tilde{\alpha}^{(2)}$, the heat capacity $C$ versus
horizon radius $r_+$ is plotted with $\tilde{\alpha}=12$ in Fig.10. The graph shows that $C$
starts from zero, for small $r_+$, $C$ is negative and gradually blows up at $r_{\tilde{c}1}$
and then becomes positive. Again $C$ vanishes at $r_+=\sqrt{\tilde{\alpha}}$ and becomes
negative. Then it blows up at $r_{\tilde{c}2}$ and change sign so that becomes positive. The black
holes are local stable for $r_{\tilde{c}1}<r_+<\sqrt{\tilde{\alpha}}$ and $r_+>r_{\tilde{c}2}$,
but are unstable in the domain $r_+<r_{\tilde{c}1}$ and $\sqrt{\tilde{\alpha}}<r_+<r_{\tilde{c}2}$.

For $D\geq9$, the temperature is always positive and there don't exist extremal black hole.
In Fig~11, heat capacity $C$ demonstrates that for small $r_+$, $C$ is negative and gradually
blows up at $r_+=r_{c_*1}$ (say) and then change sign so that $C$ becomes positive. Then,
it vanishes at $r_+=\sqrt{\tilde{\alpha}}$ and becomes negative. Again it blows up at $r_+=r_{c_*2}$ and then
becomes positive. Hence, in the region $0<r_+<r_{c_*1}$ and $\sqrt{\tilde{\alpha}}<r_+<r_{c_*2}$, black holes are
locally unstable. For $r_{c_*1}<r_+<\sqrt{\tilde{\alpha}}$ and $r_{c_*2}<r_+$, black holes are locally stable.
If taking $\tilde{\alpha}=0.1$ and $D=9$, we obtain $r_{c_*1}\approx0.18$, $\sqrt{\tilde{\alpha}}\approx0.31$
and $r_{c_*2}\approx2.90$.

%%%%%%%%%%%%%%%%%%%%%%%%%%%%%%%%%%%%%%%%%%%%%%%%%%%%%%%%%%%%%%%%%%%%%%%%%%%
\begin{figure}[htb]
\centering
\subfigure[$0.2<r_+<0.6$]{
\label{fig:subfig:a} %% label for first subfigure
\includegraphics{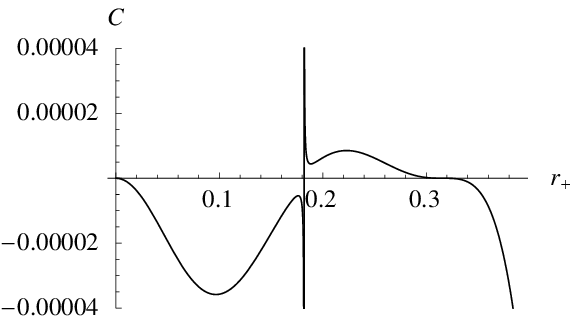}}%
\hfill%
\subfigure[$0.6<r_+<3.9$]{
\label{fig:subfig:b} %% label for second subfigure
\includegraphics{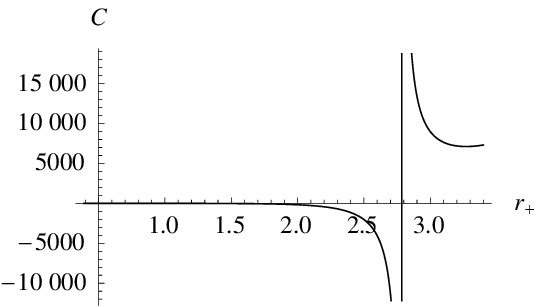}}
\caption{Heat capacity $C$ versus horizon radius $r_+$ with $l^2=10$, $D=9$ and
$\tilde{\alpha}=0.1$}
\label{fig:subfig} %% label for entire figure
\end{figure}

\section{Conclusions}
\label{sec:4}
\indent
By considering the coefficient $\hat{\alpha}_2=\alpha$ and $\hat{\alpha}_3=\alpha^2/3$,
we studied the case $\hat{\alpha}_2=\alpha<0$ and $\hat{\alpha}_3=\alpha^2/3>0$
and then presented the asymptotically AdS black hole solutions in third order Lovelock
gravity. Later, we discussed the thermodynamic properties of black holes including
gravitational mass, Hawking temperature and entropy of black holes and performed
the stability analysis of these topological black holes.

For $k=0$, all the thermodynamic and conserved quantities of the black holes don't depend on
the Lovelock coefficients and are the same as those of black holes in Einstein gravity
although the two black hole solutions are quit different.
For the horizon is negative constant hypersurface, the thermodynamics of the black
holes with Gauss-Bonnet and third order Lovelock terms are qualitatively similar to
those of black holes without these higher derivative terms. The third order Lovelock black holes
are thermodynamically stable for the whole range $r_+$.
For the positive constant hypersurface horizon, when $D=7$, there exist extremal black
holes and we found that the black hole has an intermediate unstable phase.
In eight dimensional spacetimes, however, a new phase of thermodynamically unstable small
black holes appears if the coefficient $\tilde{\alpha}$ is under a critical value.
Simultaneously, the extremal black holes also vanishes. For $D\geq 9$, black holes have similar the
distributions of thermodynamically stable regions to the case $D=8$ when the coefficient
$\tilde{\alpha}$ is under a critical value.

\end{document}